\documentclass{jaa}
\usepackage{natbib}

\usepackage{graphicx}
\usepackage{amsmath}
\usepackage{url}
\usepackage{stfloats} 
\usepackage[pdfpagelabels=false]{hyperref}	
\hypersetup{colorlinks=true,linkcolor=blue,citecolor=blue,filecolor=blue,urlcolor=blue,}
\usepackage{appendix}
\usepackage{float}
\usepackage{ulem}
\usepackage{textcmds}

\begin{document}\sloppy

\title{A Multi-frequency Study of the Candidate Double-Double Radio Galaxy J2349-0003 with a Possible Misalignment}


\author{Sudheesh T. P.\textsuperscript{1,*}, Ruta Kale\textsuperscript{2}, V. Jithesh\textsuperscript{1}, C. H. Ishwara-Chandra\textsuperscript{2}, and Joe Jacob\textsuperscript{3}}
\affilOne{\textsuperscript{1}Department of Physics and Electronics, Christ University, Hosur Main Road, Bengaluru 560029, India.\\}
\affilTwo{\textsuperscript{2}National Centre for Radio Astrophysics, TIFR, Post Bag No. 3, Ganeshkhind, Pune 411007, India.\\}
\affilThree{\textsuperscript{3}Newman College, Thodupuzha, Kerala 685584, India.}


\twocolumn[{

\maketitle

\corres{sudheesh.tp@res.christuniversity.in}

\msinfo{31 May 2025}{11 July 2025}

\begin{abstract}
We present a multi-frequency analysis of the candidate double-double radio galaxy (DDRG) J2349-0003, exhibiting a possible lobe misalignment. High-resolution uGMRT observations at Bands 3 and 4 reveal a complex radio morphology featuring a pair of inner and outer lobes, and the radio core, while the Band 5 image detects the core and the compact components. The positioning of both pairs of lobes with the central core supports its classification as a DDRG. Spectral age estimates for the inner and outer lobes indicate two distinct episodes of active galactic nucleus (AGN) activity interspaced by a short quiescent phase. The possible compact steep spectrum nature of the core, together with its concave spectral curvature, suggests ongoing or recent jet activity, suggesting the possibility that J2349–0003 may be a candidate triple-double radio galaxy. With a projected linear size of 1.08 Mpc, J2349-0003 is classified as a giant radio galaxy (GRG), although its moderate radio power ($\sim$$10^{24} \mathrm{WHz}^{-1}$) suggests a sparse surrounding environment. Arm-length (R$_\theta$) and flux density ratios (R$_S$) indicates environmental influences on source symmetry. The observed lobe misalignment and the presence of nearby galaxies in the optical image suggest that merger-driven processes may have played a key role in shaping the source's evolution.

\end{abstract}

\keywords{radio continuum: galaxies – galaxies: active.}

}]


\doinum{12.3456/s78910-011-012-3}
\artcitid{\#\#\#\#}
\volnum{000}
\year{0000}
\pgrange{1--}
\setcounter{page}{1}
\lp{1}

\section{Introduction}
Restarted radio galaxies are a class of active galaxies that often exhibit evidence of multiple episodes of jet activity, driven by the central active galactic nucleus (AGN). Such episodic AGN activity suggests that the accretion process onto the central super massive black hole (SMBH) is not continuous but occurs in distinct phases, leading to the formation of multiple pairs of radio lobes, observed in some radio galaxies. Such radio galaxies which are characterised by two pairs of radio lobes are known as double-double radio galaxies \citep[DDRG;][]{2000schoenmakers1}. Various radio properties of this class of sources have been extensively studied by multiple authors \citep[e.g.][]{2006saikia, 2006konar, 2011joshi, 2019mahatma, 2019nandi}. Sources with more than two pairs of radio lobes are also identified. Such sources with three episodes of jet activity are known as triple-double radio galaxies \citep{2007brocksopp}.

Studies on episodic radio sources reveal that intermittent AGN activity is not exclusively characterised by the presence of a distinct \lq{double-double}\rq morphology. An example of such a source is 3C388. Spectral index studies of the radio source 3C388 by \citet{1994roettiger} identified two distinct emission regions—one with a flat spectral index and the other with a steep spectral index—separated by a well-defined transition layer. They interpreted this as evidence of two distinct epochs of jet activity. Similarly, \citet{2022timmerman} observed ring-like structures within the lobes of the Hercules A radio galaxy, which they attributed to intermittent AGN activity. Recently, \citet{2025raj} proposed a method to identify restarted radio galaxies that do not exhibit clear morphological evidence of episodic jet activity. Nevertheless, DDRGs remain the most readily identifiable class of restarted sources based on their morphology.

\citet{2019mahatma} suggest that DDRGs are characterised by two pairs of edge-brightened radio lobes. However, this is not always the case. The outer lobes can appear diffuse, lacking compact hotspots, while the inner lobes may contain compact peaks, indicating ongoing jet activity. Since the outer lobes consist of an aging electron population with energy loss, their spectrum will be steeper compared to the inner lobes, which are still being fed with energetic particles by the jets \citep[e.g.][]{2000schoenmakers2}. 

The morphology of DDRGs is often closely linked to their surrounding environments. The initial jets from the central AGN propagate through the intergalactic medium (IGM), expanding into a diffuse cocoon. After a quiescent phase, the restarted jets traverse this pre-existing cocoon, which possesses distinct pressure and density characteristics compared to the IGM, leading to altered propagation dynamics \citep{2024dabhade}. In addition, intrinsic AGN properties such as accretion rate and jet power, also play a critical role in shaping the morphology of DDRGs \citep{2014walg, 2020walg, 2021nandi}. Thus, the formation of DDRGs is a complex process influenced by both AGN properties and the ambient environment.

Various theories have been proposed in the literature to explain the formation of DDRGs. Simulations by \citet{1991clarke} and \citet{2014walg}, along with analytical studies by \citet{2000kaiser}, explore different possible mechanisms for the origin and evolution of episodic radio sources, including DDRGs. Another scenario was discussed by \citet{2003liu}, who proposed that galaxy mergers, leading to the formation of binary black hole systems, could trigger episodic AGN activity, thereby resulting in the formation of DDRGs. However, a universally accepted and well-tested model is still lacking, because the formation of DDRGs are due to a complex interplay between intrinsic AGN properties and environmental conditions. Furthermore, most of the observed DDRGs tend to be large and luminous, which may be the result of the selection bias. Earlier studies—such as those by \citet{2012nandi} and \citet{2017kuzmicz}—relied on data from surveys like the NRAO VLA Sky Survey \citep[NVSS;][]{1998condon}, and the Faint Images of the Radio-Sky at Twenty centimetres \citep[FIRST;][]{1995becker}, which are less sensitive to low-luminosity sources. More recently, low-frequency observations from the LOFAR Two-Metre Sky Survey \citep[LoTSS;][]{2017shimwell}, as used by \citet{2019mahatma} and \citet{2024dabhade}, along with GMRT studies by \citet{2019nandi}, have identified DDRGs with lower luminosities and sizes ranging from a few hundred kiloparsecs to over 1 Mpc. 

Although in most DDRGs, the outer lobes are reasonably well aligned with the inner lobes, there are cases where a pronounced misalignment exists between the two. A notable example is the DDRG 3C 293, where the inner double is misaligned with respect to the outer lobes by $\sim$35\textdegree\ \citep{2011joshi, 2022kukreti}. Such misalignments are rare in radio galaxies, and a few DDRGs with extreme misalignments were recently identified by \citet{2024dabhade}. In the case of 3C 293, \citet{1981bridle} proposed that the observed misalignment might result from the refraction of radio jets caused by pressure gradients within the dense circumgalactic medium, while \citet{2016machalski} suggested that it might be caused by a fast reorientation of the jet axis, potentially triggered by a rapid change in the spin direction of the central black hole. A notable observation on the formation of misaligned DDRG was made by \citet{2017nandi}. The authors found evidence for a binary black hole at the core of the misaligned DDRG J1328+2752. However, the formation of these DDRGs with pronounced misalignment is still a matter of debate.

This paper discusses the properties of a candidate DDRG J2349–0003 (RA: 23h49m29.77s, DEC:-00d03m05.87s) identified from the Galaxy Evolution and Magnetisation in the Saraswati Supercluster (GEMSS) survey (Kale et al., in preparation). GEMSS is a dedicated radio sky survey (PI: Ruta Kale) focused on the Saraswati supercluster region \citep{2017bagchi}, and is being conducted using Band 3 of the upgraded Giant Metrewave Radio Telescope (uGMRT). During our analysis of the GEMSS radio maps, we discovered this potential DDRG candidate showing signs of episodic activity and possible lobe misalignment, which has not yet been explored in detail. Although the source was previously reported as a Giant Radio Galaxy (GRG) by \citet{2017dabhade}, with a projected linear size of 0.84 Mpc, its complex morphology is revealed in this study for the first time. A misaligned Giant Double-Double Radio Galaxy (GDDRG) can serve as a valuable tracer for understanding the formation mechanisms of episodic radio activity, as well as a probe of the ambient environment in which it evolves. In this paper, we present deep, multi-frequency radio observations of this candidate GDDRG with potential misalignment, using uGMRT.

The structure of this paper is as follows. Section~\ref{sec:observations} provides details on the uGMRT observations and the data analysis of J2349-0003. Section~\ref{sec:results} presents the radio properties of the source together with the results obtained. Section~\ref{sec:discussions} provides a detailed discussion. Section~\ref{sec:conclusion} summarises our findings and conclusions.

In this study, we adopted a flat $\Lambda$CDM cosmological model based on the results from the Planck Collaboration (H\,$_0$ = 67.8 km\,s$^{-1}$\,Mpc$^{-1}$, $\Omega_{m}$ = 0.308 and $\Omega_{\Lambda}$ = 0.692; \citep{2016planck}). At $z$ = 0.187, this corresponds to a scale of 3.227 kpc/". We define spectral index $\alpha$ such that \,S$_{\nu}$ $\propto$ $\nu^{\alpha}$, where \,S$_{\nu}$ is the flux density at a given frequency $\nu$.

\section{Observations and Data Analysis}
\label{sec:observations}
J2349-0003 was observed using the upgraded Giant Metrewave Radio Telescope (uGMRT), which offers nearly continuous frequency coverage from 50 to 1500 MHz \citep{2017gupta}. uGMRT operates in four main observing bands-Band 2 (centered at 200 MHz), Band 3 (centered at 400 MHz), Band 4 (centered at 650 MHz), and Band 5 (centered at 1250 MHz). Our observations were conducted in Bands 4 and 5 (Project Code: 43\_125, PI: Sudheesh T P) with the GMRT Wideband Backend \citep[GWB;][]{2017reddy}, with 200 MHz bandwidth divided into 2048 spectral channels. The Band 4 observation took place on October 31, 2022, whereas the Band 5 observation was conducted on November 05, 2022. We have summarised the details of the observation in Table ~\ref{tab:ugmrt-obs-summary}.

\par The uGMRT data were processed using the {\tt CAPTURE} \footnote{\url{https://github.com/ruta-k/CAPTURE-CASA6}} pipeline \citep{2021kale}, a continuum imaging pipeline that uses Common Astronomy Software Applications \citep[{\tt CASA}; ][]{2007mcmullin} to process the data. Initially, the data were flagged to remove bad antennas and mitigate RFI. The flux density of the primary calibrators was set based on the Perley–Butler 2017 flux scale \citep{2017perley}. Following standard calibration procedures, the target source data were calibrated. The calibrated data were then split and subsequent flagging was performed using different flagging modes. To manage data volume while avoiding bandwidth smearing, 10 frequency channels were averaged. These processes were applied consistently to both Bands 4 and 5. Imaging of the target visibilities was carried out using the {\tt CASA} task {\tt tclean}. The images were produced with ‘{\tt briggs}’ weights and {\tt robust} = 0, which provides a balance between resolution and surface brightness sensitivity. During imaging and self-calibration, the target source measurement set was divided into 8 frequency sub-bands, each containing 20 channels, effectively splitting the total 200 MHz bandwidth into 8 spectral windows. Imaging and self-calibration were performed on these sub-banded data files. To perform the primary beam correction of the images, we have used the task {\tt ugmrtpb} \footnote{\url{https://github.com/ruta-k/uGMRTprimarybeam}}.

The source also contains Band 3 uGMRT data (Project Code: 34\_059, PI: Ruta Kale). It is observed as a part of the \qq{Galaxy Evolution and Magnetization in the Saraswati Supercluster} (GEMSS) survey. The data were recorded using GMRT Software Backend \citep[GSB;][]{2010roy}, across a bandwidth of 33.3 MHz, divided into 256 spectral channels. The data were reduced using the {\tt CAPTURE} pipeline, following standard flagging, calibration, imaging and self-calibration procedures. Band 3 image was also produced with ‘{\tt briggs}’ weights and {\tt robust} = 0. However, due to the limited bandwidth of the GSB data, sub-banding was not applied, unlike in Bands 4 and 5. The observation details of Band 3 data are also provided in Table ~\ref{tab:ugmrt-obs-summary}. 
 
 \begin{table*}[h]
	\centering
	\caption{Summary of uGMRT observations of J2349-0003.}
	\label{tab:ugmrt-obs-summary}
	\begin{tabular}{lccc} 
		\hline
		   & Band 3 & Band 4 & Band 5\\
		\hline
		Frequency range (MHz) & 300-500 & 550-750 & 1000-1450\\
		Number of channels & 256 & 2048 & 2048\\
		Bandwidth (MHz) & 33.3 & 200 & 200\\
            On source time (min) & 120 & 432 & 450\\
            Integration time (sec) & 16 & 8 & 8\\
            Flux calibrator & 3C48 & 3C48 & 3C48\\
            Phase calibrator & 2225-049 & 2225-049 & 0022+002\\
		\hline
	\end{tabular}
\end{table*}

\section{Results}
\label{sec:results}
\begin{figure*}
    \centering
    \includegraphics[width=5.5in, height = 2.5in]{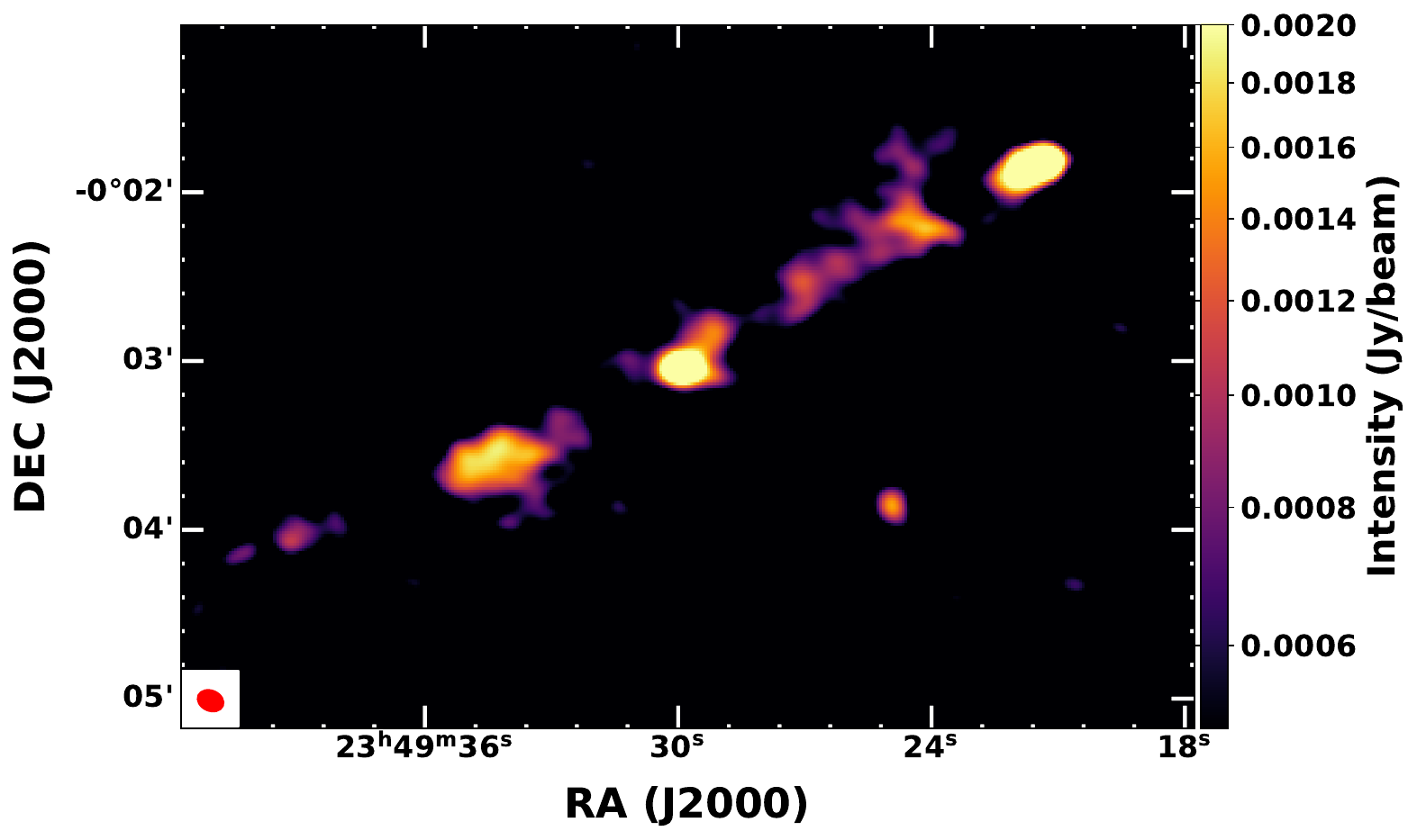}
    \includegraphics[width=5.5in, height = 2.5in]{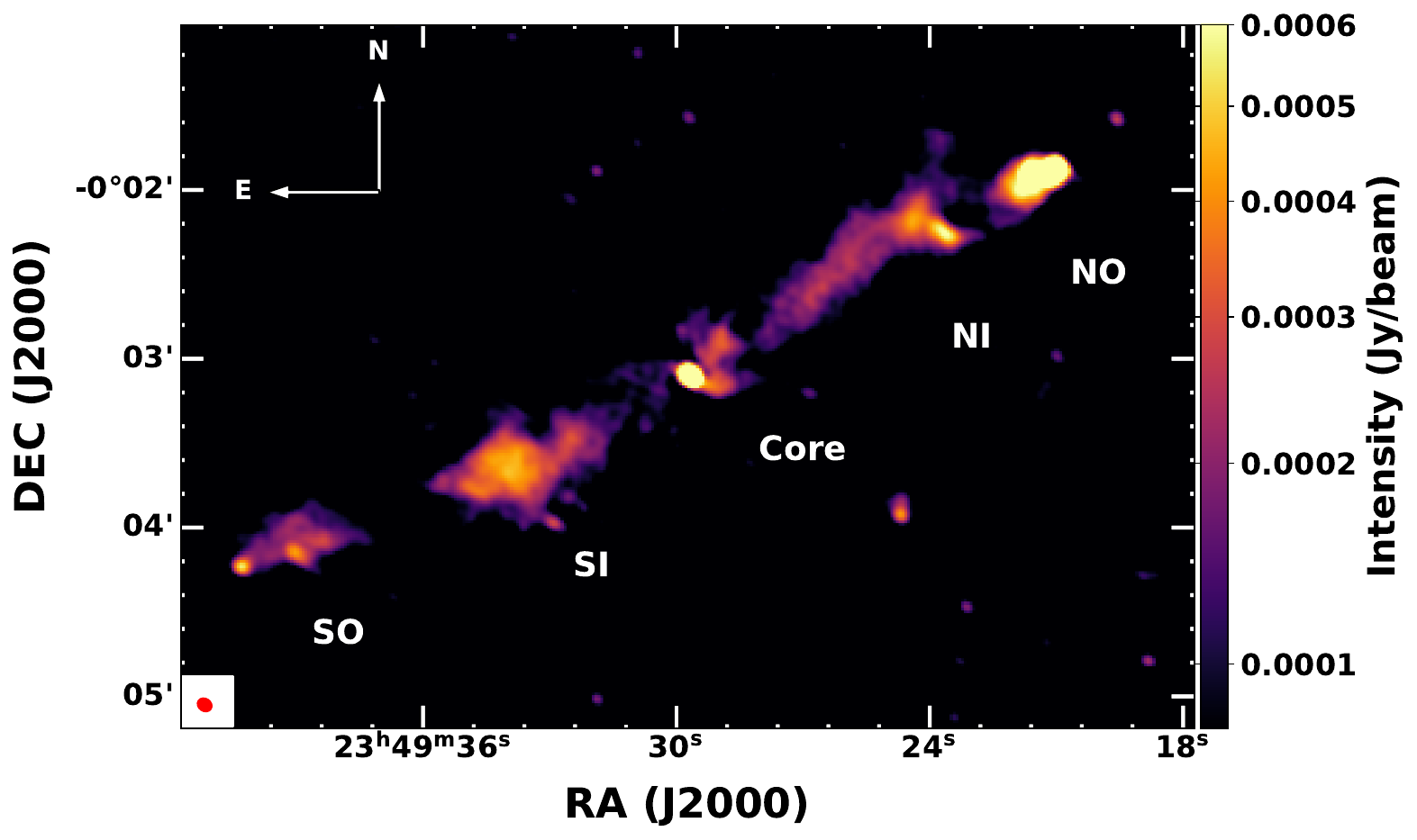}
    \includegraphics[width=5.5in, height = 2.5in]{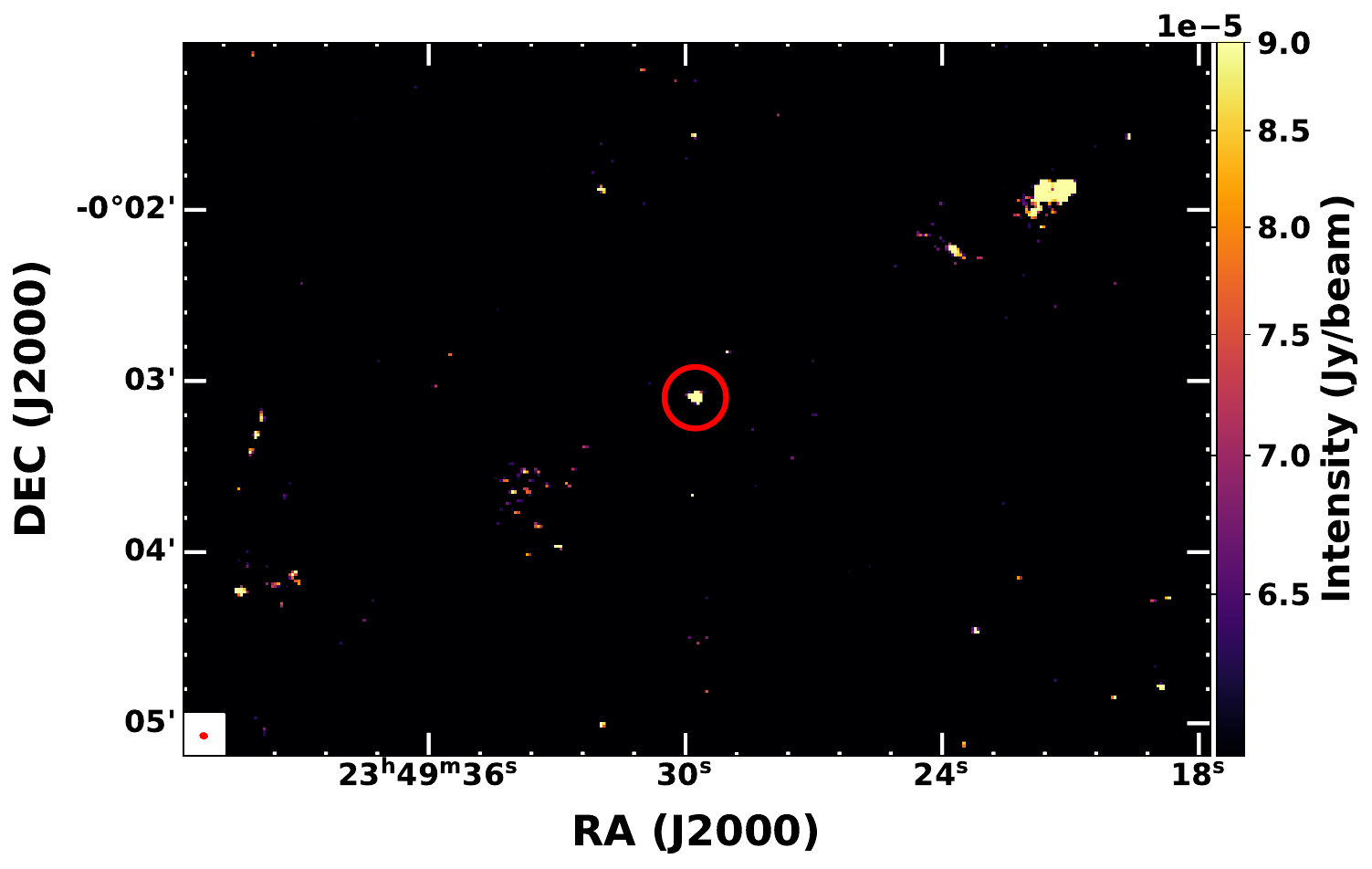}
    \caption{uGMRT radio images of J2349-0003 at Bands 3 (top), 4 (middle), and 5 (bottom). The middle panel contains the labels of various radio components of J2349-0003: southern outer lobe (SO), southern inner lobe (SI), northern inner lobe (NI), northern outer lobe (NO), and core. The radio beams are shown at the bottom left corner of the images. The high resolution Band 5 image shows the radio core (red circle) and the compact components in the two pairs of lobes.}
    \label{fig:DDRG-highres-images}
\end{figure*}

J2349-0003 is hosted by the optical galaxy SDSS J234929.77-000305.8, at a photometric redshift ($z$) of 0.187$\pm$0.0491. Using radio observations with uGMRT, we obtained high-resolution, deep-radio maps of J2349-0003 in Bands 3, 4, and 5, which are given in Fig.~\ref{fig:DDRG-highres-images}. The information on the images is provided in Table~\ref{tab:b3-b4-b5-image_details}. 

\begin{table*}
	\centering
	\caption{Details of the Band 3, 4, and 5 images of J2349-0003 shown in Figure ~\ref{fig:DDRG-highres-images}.}
	\label{tab:b3-b4-b5-image_details}
	\begin{tabular}{lccc}
		\hline
		Band & rms & Resolution & PA\\
             & (mJy/beam) & (arcsec $\times$ arcsec) & (degree)\\
		\hline
		Band 3 (300-500 MHz) & 0.167 & 9.5" $\times$ 7.2" & 66.6\\
		Band 4 (550-750 MHz) & 0.025 & 5.2" $\times$ 4.4" & 56.0\\
            Band 5 (1050-1450 MHz) & 0.019 & 2.4" $\times$ 1.9" & 80.4\\
		\hline
	\end{tabular}
\end{table*}

\subsection{Radio Morphology}
\begin{figure*}
    \centering
    \includegraphics[width=\textwidth]{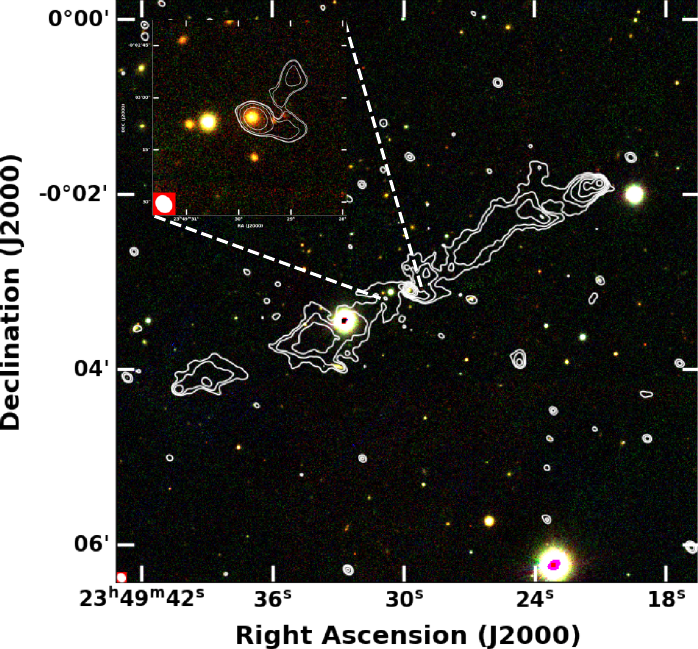}
    \caption{uGMRT Band 4 radio contours overlaid on a Pan-STARRS color composite image (combined g, r, and i filters). Six radio contours at levels of 3$\sigma \times$ [1, 2, 4, 8, 16, 32] are shown, where $\sigma$ = 25 $\mu$Jy/beam represents the rms noise around the source. The bottom left corner shows the radio beam. The bright optical source seen in the southern inner lobe is the foreground star Gaia DR2 2642294499322549376. The inset at the top left corner shows the position of the host galaxy. Six equally spaced Band 4 contours (in log scale) starting from 3$\sigma$ are overlaid over the host galaxy in the inset, shows the ejection of radio jet.}
    \label{fig:optical-radio-overlay}
\end{figure*}

The high-resolution uGMRT Band 3, 4, and 5 radio images reveal the detailed morphology of J2349-0003. A radio core, coincident with the optical galaxy SDSS J234929.77-000305.8, is clearly detected in Band 3, 4, and 5 radio images. The Band 3 image shows a central region, two asymmetric inner lobe-like structures without compact hotspots, and two outer lobe-like structures. Based on this morphology, the radio components are identified as the core, northern outer lobe (NO), northern inner lobe (NI), southern inner lobe (SI), and southern outer lobe (SO), as illustrated in the middle panel of Fig.~\ref{fig:DDRG-highres-images}. The high-resolution Band 4 map shows two compact hotspot-like features in the southern outer lobe and a knot/kink-like feature in the northern inner lobe. Additionally, the Band 4 image indicates that the radio emission from the central region is initially ejected towards the west and east, before bending toward the north and south, respectively. The radio maps also suggest a possible misalignment of the radio lobes, which could either be intrinsic or a result of projection effects. Fig.~\ref{fig:optical-radio-overlay}. represents the Band 4 radio contours overlaid over Pan-STARRS \citep{2016chambers} color composite image. The figure shows the various radio components associated to a single host galaxy. The inset shows the direction of emission of radio jet from the core. No optical counterparts have been identified in any of the compact radio components outside the core region, supporting the likelihood that both pairs of lobes are associated with a single central core, making J2349-0003 a strong DDRG candidate. In the following sections, the episodic nature of the source and the origin of this misalignment will be discussed in detail. 

\subsection{Linear Size}
The Band 3 radio map, which is the most sensitive to diffuse emission among our data, is used to estimate the maximum extent of the source. Based on this map, the largest angular size (LAS) is measured to be 5.56$^\prime$, determined at the lowest contour level of 3$\sigma$, where $\sigma$ = 0.167 mJy beam$^{-1}$ represents the local rms noise around the source in Band 3. The projected linear size is then calculated using the relation:

\begin{equation}
    D=\frac{\theta \times D_{c}}{(1+z)} \times \frac{\pi}{10800}
	\label{eq:linsize}
\end{equation}

\noindent where D is the projected linear size of the radio source in Mpc, $\theta$ is the angular size in arcmin, $D_{c}$ is the comoving distance in Mpc and $z$ is the redshift of the host galaxy. Using the relation, the projected linear size is estimated to be 1.08$\pm$0.29 Mpc. This places the source in the class of Giant Radio Galaxies (GRGs) which are defined to be radio galaxies with a projected linear size greater than 0.7 Mpc \citep[e.g.][]{2020dabhade}.

\subsection{Flux Density and Radio Power}
\label{sec:fluxdensity}
The integrated flux densities of J2349-0003 in Bands 3 and 4 were determined using the {\tt imview} task in {\tt CASA}. To calculate the flux density, we manually selected emission regions exceeding the 3$\sigma$ threshold in the radio maps of these bands. The band 5 radio map primarily reveals compact components, including the core. We used the {\tt CASA} task {\tt imfit} to measure the core flux density in this band. The uncertainty in the flux density of extended sources was determined using the relation:

\begin{equation}
    \delta S = \sqrt{\left(S_{\nu}\times \delta S_{c}\right)^2+N_{beam}\times\left(\sigma\right)^2}
    \label{eq:fluxerr}
\end{equation}

\noindent where $\delta S$ is the uncertainty in flux density, $S_{\nu}$ is the flux density measured at frequency $\nu$, $\delta S_{c}$ is the calibration error, $\sigma$ is the rms noise of the radio map, $N_{beam}$ is the number of beams. For Bands 3, 4, and 5, the calibration error $\delta S_{c}$ is taken as 10\% \citep{2017chandra, 2025kale}. The Band 5 radio map shows only the compact components; therefore, we did not measure the integrated flux density from this high-resolution map. The integrated flux densities measured in the different bands are listed in Table~\ref{tab:basic_radio_parameters}. The radio power of the source was estimated using the relation:

\begin{equation}
    P_{\nu} = 4\pi D_{L}^2S_{\nu}(1+z)^{-\alpha -1}
    \label{eq:radpower}
\end{equation}

\noindent where $P_{\nu}$ is the radio power, $D_{L}$ is the luminosity distance at the redshift $z$, $S_{\nu}$ is the flux density and $(1+z)^{-\alpha -1}$ is the standard k correction term where $\alpha$ is the radio spectral index. The estimated radio powers in Bands 3 and 4 are also provided in Table~\ref{tab:basic_radio_parameters}.

\begin{table}
	\centering
	\caption{Basic radio parameters of J2349-0003.}
	\label{tab:basic_radio_parameters}
	\begin{tabular}{lc}
		\hline
		Properties & Values\\
		\hline
		Redshift ($z$) & 0.187 $\pm$ 0.0491 \\
            LAS ($^\prime$) & 5.56 \\
            Projected Linear Size (Mpc) & 1.08 $\pm$ 0.29 \\
            S$_{322 MHz}$ (mJy) & 90.8 $\pm$ 9.3 \\
            S$_{648 MHz}$ (mJy) & 65.9 $\pm$ 6.6 \\
            P$_{322 MHz}$ ($\times 10^{24} W\,\mathrm{Hz}^{-1}$) & 9.3 $\pm$ 4.6 \\
            P$_{648 MHz}$ ($\times 10^{24} W\,\mathrm{Hz}^{-1}$) & 6.7 $\pm$ 3.3 \\
		\hline
	\end{tabular}
\end{table}

\subsection{Spectral Index}
\begin{figure}
    \includegraphics[width=\columnwidth]{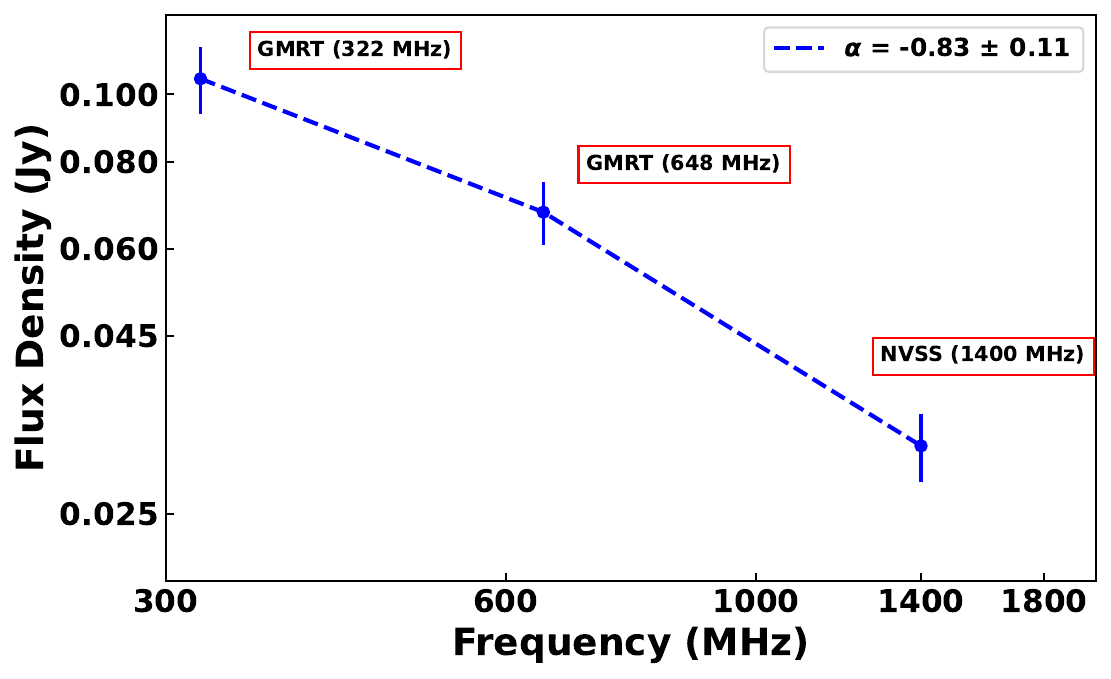}
    \caption{Integrated spectrum of J2349-0003 using uGMRT Band 3, 4, and NVSS data.}
    \label{fig:integrated_spectrum}
\end{figure}

Along with our uGMRT Band 3 and Band 4 observations, J2349-0003 has also been studied in the NRAO VLA Sky Survey \citep[NVSS; ][]{1998condon}. NVSS operates at a frequency similar to our Band 5 data; however, it is more sensitive to extended emission, whereas the high-resolution Band 5 data does not capture such structures. Therefore, we used the NVSS radio map to estimate the 1400 MHz flux density. The NVSS radio map of J2349–0003 was retrieved from the NVSS archive, and its flux density was determined as described in Section~\ref{sec:fluxdensity}. For uniformity, the Band 3 and Band 4 data were convolved to match the resolution of the NVSS map before estimating their integrated flux densities. Further, the same regions were used to estimate the flux densities from all three radio maps. Using the same method, we also estimated the Band 5 flux density, which is found to be 18.17$\pm$2.4 mJy. This value is relatively low compared to the NVSS flux density, possibly due to the lower sensitivity of Band 5 to extended diffuse emission, when compared to NVSS. Therefore, we used the NVSS flux density in our integrated spectrum. A summary of the observed spectrum of J2349-0003 is presented in Table~\ref{tab:observed_spectrum_b3_b4_nvss}, while its integrated spectrum is shown in Fig.~\ref{fig:integrated_spectrum}. The observed spectrum follows a single power-law fit within the 300–1500 MHz frequency range, yielding an integrated radio spectral index of -0.83$\pm$0.11.

\begin{table}
	\centering
	\caption{Observed spectrum of J2349-0003 using Band 3, 4, and NVSS radio maps.}
	\label{tab:observed_spectrum_b3_b4_nvss}
	\begin{tabular}{lcc}
		\hline
		Data & rms & Flux density \\
             & (mJy/beam) & (mJy) \\
		\hline
		Band 3 (322 MHz) & 1.13 & 105.3 $\pm$ 11.5\\
		Band 4 (648 MHz) & 0.462 & 67.7 $\pm$ 7.0\\
            NVSS (1400 MHz) & 0.556 & 31.3 $\pm$ 3.5\\
		\hline
	\end{tabular}
\end{table}

\begin{figure*}
    \includegraphics[width=\textwidth]{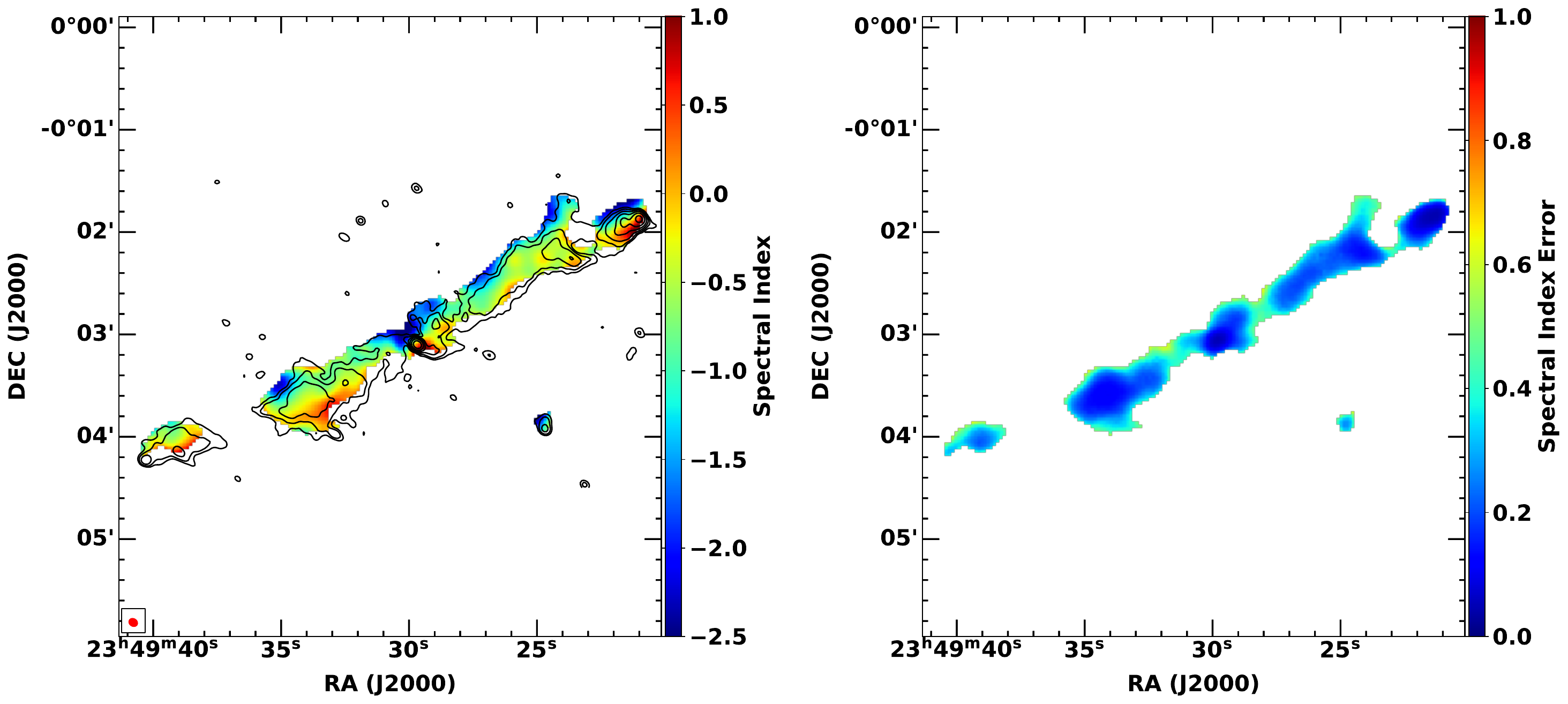}
    \caption{Left panel: Spectral index map using uGMRT Band 3 and Band 4 radio maps. The uGMRT Band 4 contours are overlaid with six levels in multiples of 2, starting from 3$\sigma$ where $\sigma$ = 0.025 mJy/beam. The bottom left corner shows the radio beam. Right panel: Spectral index error map.}
    \label{fig:spectral_index_map}
\end{figure*}

We also generated a spatially resolved spectral index map of the source using data from Band 3 and Band 4. Such maps are particularly useful for analysing the physical mechanisms shaping the observed morphology of J2349-0003. To enhance the detection of diffuse emission, the Band 3 and Band 4 radio maps were re-imaged using the same UV range of 0-29k$\lambda$, applying the {\tt briggs} weighting scheme with {\tt robust} = +0.5. The maps were then convolved to a uniform resolution of 10$^{\prime\prime} \times$10$^{\prime\prime}$ and regridded to a common pixel scale. To ensure reliable spectral index calculations, only pixels with flux values exceeding 3$\sigma$, where $\sigma$ represents the local rms noise in the respective radio map, were considered. These radio maps were then combined to generate the spectral index map and its corresponding error map, as shown in Fig.~\ref{fig:spectral_index_map}. The spectral index map shows relatively steep spectral indices ($\alpha < -0.5$) throughout the source. If the inner lobes were being powered by the current epoch of jet activity, they would be expected to appear flatter, an effect that is not clearly evident in the resolved spectral index map.

\begin{figure*}
    \includegraphics[width=\textwidth]{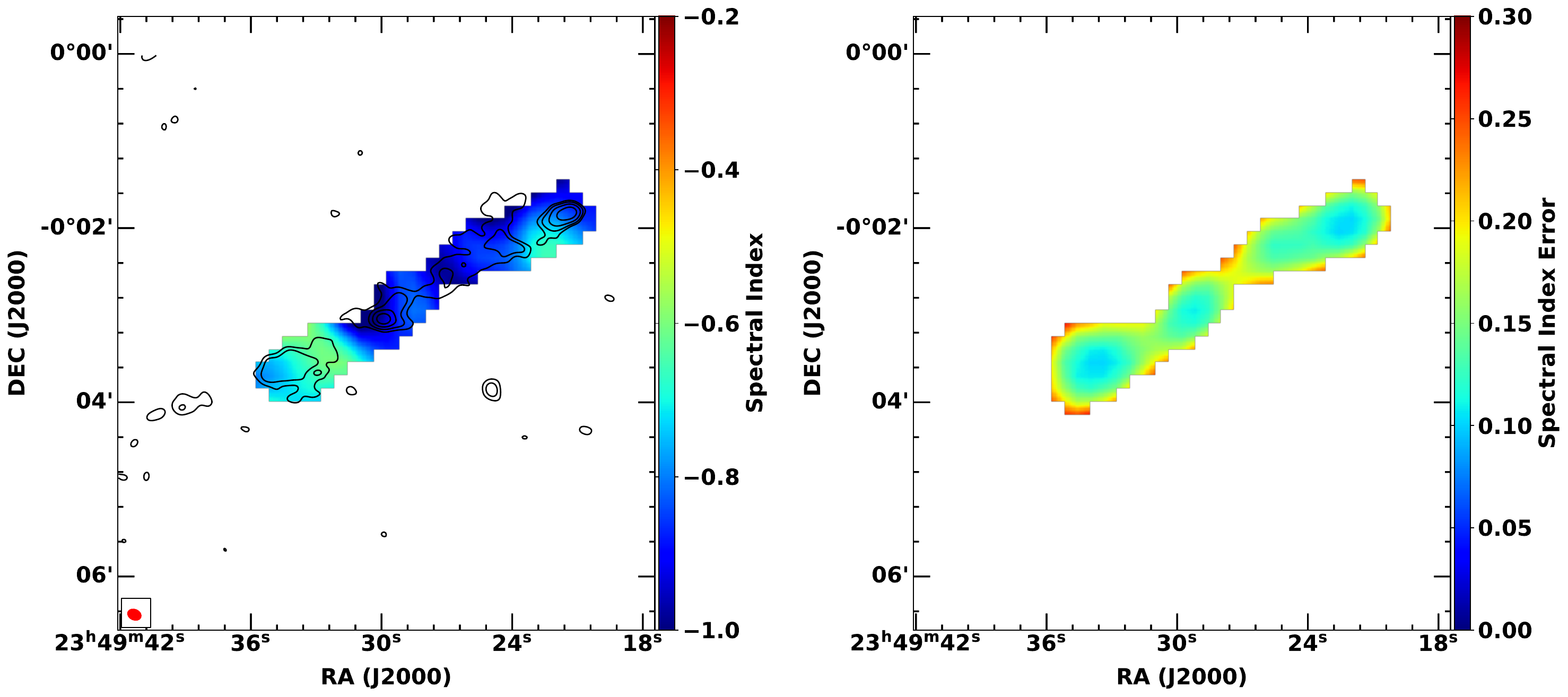}
    \caption{Left panel: Spectral index map using uGMRT Band 3 and NVSS radio maps. The uGMRT Band 3 contours are overlaid with six levels in multiples of 2, starting from 3$\sigma$ where $\sigma$ = 0.167 mJy/beam. The bottom left corner shows the radio beam. Right panel: Spectral index error map.}
    \label{fig:spectral_index_map-b3-nvss}
\end{figure*}

In order to get a broader picture about the intrinsic AGN properties and the environmental factors that are shaping the source morphology, we also generated a spectral index map using Band 3 and NVSS radio maps. The Band 3 map was convolved to the NVSS resolution of 45$^{\prime\prime} \times$45$^{\prime\prime}$ and both were regridded to a common pixel scale. To ensure a reliable signal-to-noise ratio (SNR), only pixels with flux values exceeding 3$\sigma$, where $\sigma$ represents the local rms noise in the respective radio map, were considered. The generated spectral index map and its corresponding error map, are provided in Fig.~\ref{fig:spectral_index_map-b3-nvss}. This is a low-resolution map compared to Fig.~\ref{fig:spectral_index_map}, and the different radio components are not resolved. However, the spectral index across the lobes is predominantly steep, indicating an aged synchrotron plasma with minimal evidence of recent re-acceleration. The two-point spectral index between Band 3 and NVSS is calculated to be -0.83$\pm$0.11, supporting the overall steep nature of the emission. Furthermore, the northern lobes appear steeper than the southern lobe. This trend is not observed in the Band 3-Band 4 spectral index map. The observed asymmetry could be due to differences in synchrotron aging, with the northern lobes being older than the southern lobe. Alternatively, it could result from environmental effects as well, where particles in the northern lobe experience greater radiative losses due to interaction with a denser medium. Projection effects may also contribute to the observed steepness difference.

\subsection{Magnetic Field and Spectral Age}
To determine whether J2349–0003 is a DDRG, we need to estimate the spectral ages of both pairs of lobes. In the case of a DDRG, the outer lobes are expected to host an older population of electrons, which have lost more energy over time through synchrotron and inverse-Compton losses, resulting in a higher spectral age. In contrast, the inner lobes may still be actively powered by the central AGN, containing more energetic electrons and therefore exhibiting a relatively lower spectral age. Thus, comparing the spectral ages of the two pairs of lobes provides a diagnostic tool for assessing the possibility of J2349–0003 being a DDRG.

The energy losses of injected electrons manifest as a steepening of the radio spectrum beyond a characteristic frequency, known as the break frequency. Once the break frequency and the magnetic field strength of the source are determined, the spectral age can be estimated using the relation provided in \citet{2018turner} as:

\begin{equation}
    \tau = \frac{\nu B^{1/2}}{B^{2}+B_{ic}^{2}}\left[\nu_{b} (1+z)\right]^{-1/2}
    \label{eq:spectral-age}
\end{equation}

\noindent where $B$ is the magnetic field strength of the source, $\nu_{b}$ is the break frequency in GHz, $z$ is the redshift of the source, and $B_{ic}$ = 0.318$(1+z)^2$ is the magnitude of the magnetic field equivalent to the microwave background. The constant of proportionality $\nu$ is given by equation~(\ref{eq:spectral-age-constant}). 

\begin{equation}
    \nu = \left(\frac{243 \pi m_{e}^{5}c^{2}}{4\mu_{0}^{2}e^{7}}\right)^{1/2}
    \label{eq:spectral-age-constant}
\end{equation}

\noindent where $\mu_{0}$ is the magnetic permeability of free space. During the estimation, it is assumed that the magnetic field strength remains constant within the source over the considered time interval of the process. Additionally, energy losses are assumed to occur exclusively through synchrotron radiation and inverse-Compton scattering, with no other significant loss mechanisms. Under the assumption of minimum energy conditions, the minimum energy density is computed using the following relation given in \citet{2004govoni} as: 

\begin{equation}
    u_{min} = \xi (\alpha,\nu_{1},\nu_{2})(1+k)^{4/7}(\nu_{0})^{4\alpha /7}(1+z)^{(12+4\alpha) /7}(I_{0}/d)^{4/7}
    \label{eq:minenergy}
\end{equation}

\noindent Here, $u_{min}$ represents the minimum energy density, $\nu_{1}$ and $\nu_{2}$ are the lower and upper frequency limits within which the spectrum is integrated, and $k$ denotes the ratio of the energy of relativistic protons to that of electrons (1 or 100). The parameter $\xi (\alpha,\nu_{1},\nu_{2})$ is calculated by \citet{2004govoni} for different spectral index values. $I_{0}$ refers to the surface brightness in $mJyarcsec^{-2}$ measured at a frequency of $\nu_{0}$ MHz. $d$ denotes the depth of the source in kpc. For this calculation, we have assumed $\nu_{1}$ = 10 MHz, $\nu_{2}$ = 10 GHz, $k$ = 1. We utilized the lowest frequency Band 3 radio map for our estimations. Assuming a cylindrical shape for the radio source, we calculated the source depth $d$. The surface brightness $I_{0}$ was also measured from the Band 3 radio map at a frequency of $\nu_{0}$ = 322 MHz, and we assumed a filling factor of unity. With the knowledge of minimum energy density (u$_{min}$), the magnetic field can be calculated as:

\begin{equation}
    B_{eq} = \left(\frac{24\pi}{7} u_{min}\right)^{1/2}
    \label{eq:magfield-eq}
\end{equation}

Equation~\ref{eq:magfield-eq} is based on the classical formalism. \citet{2005beck} has revised the expression by taking into account of the value of minimum Lorentz factor ($\gamma_{min}$). Using a value of $\gamma_{min}$ = 100, we have calculated the magnetic field using the following modified expression:

\begin{equation}
    B_{eq}^{'} = 1.1\gamma_{min}^{\frac{1-2\alpha}{3+\alpha}}B_{eq}^{\frac{7}{2(3+\alpha)}}
    \label{eq:magfield-eqmod}
\end{equation}

\begin{figure*}
    \includegraphics[width=\columnwidth]{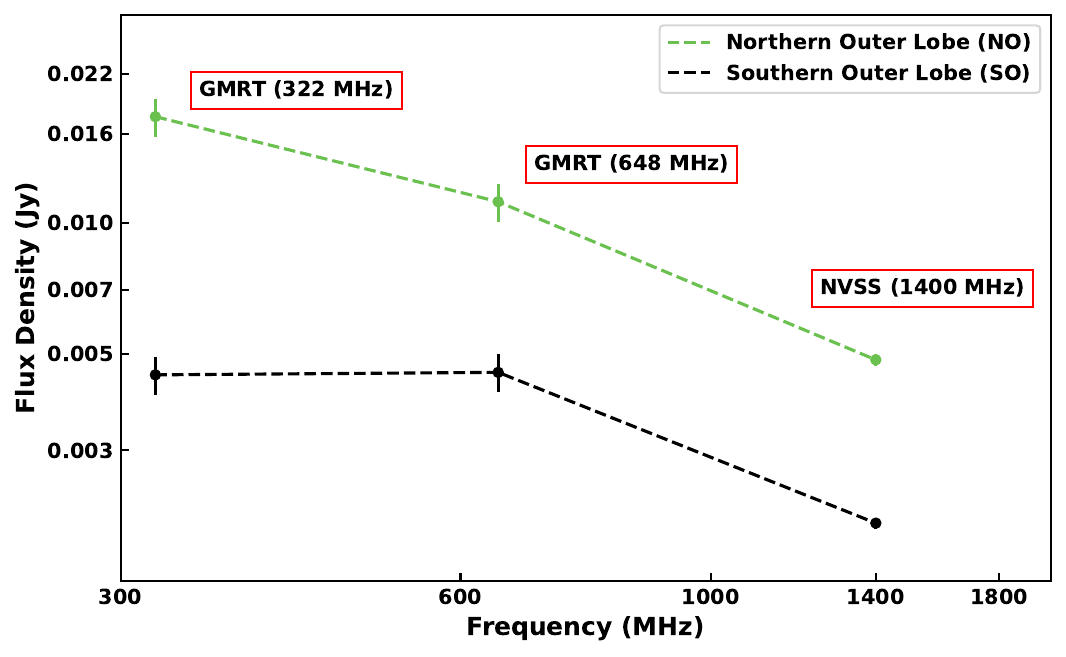}
    \includegraphics[width=\columnwidth]{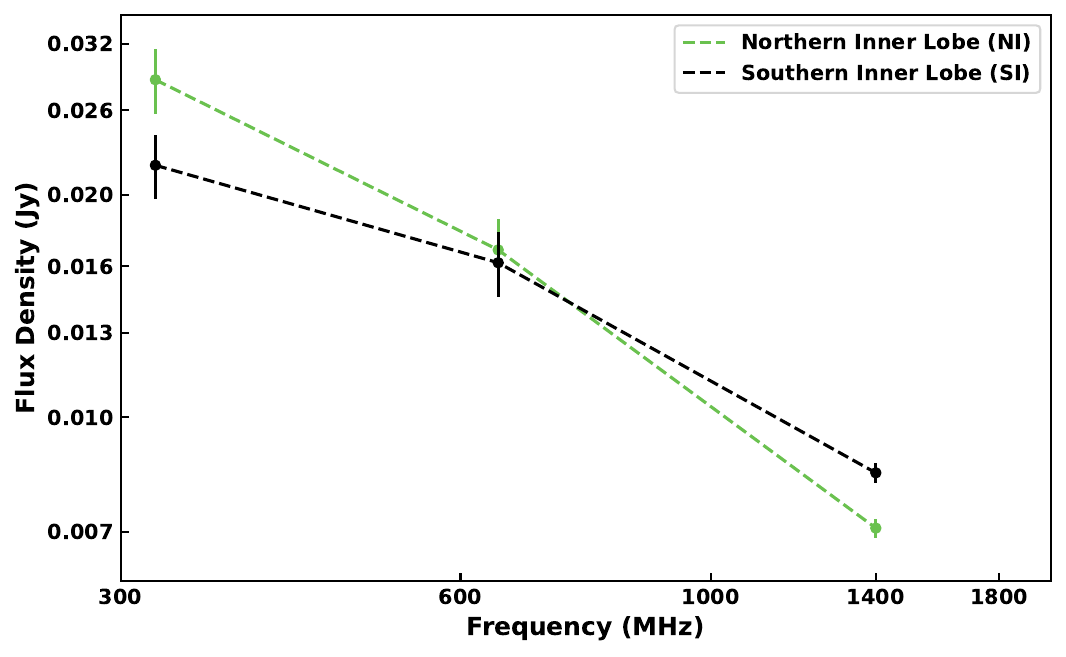}
    \caption{Left panel: Integrated spectrum of the outer lobes using uGMRT Band 3, 4, and NVSS radio data. Right panel: Integrated spectrum of the inner lobes using uGMRT Band 3, 4, and NVSS radio data. The flux densities are estimated using convolved radio maps with a common resolution.}
    \label{fig:lobe_spectrum}
\end{figure*}

In order to investigate the episodic nature of J2349-0003, we estimated the magnetic field strengths and spectral ages of both the outer and inner pairs of lobes. If these two pairs are associated with distinct epochs of AGN activity, their spectral ages are expected to differ, with the outer lobes being older than the inner ones. We plotted the integrated spectra of the outer and inner pairs of lobes, as shown in Fig.~\ref{fig:lobe_spectrum}. The spectra exhibit a possible break frequency ($\nu_{b}$) around 648 MHz. Using the estimated magnetic field strengths and the break frequency, we calculated the spectral ages of the outer and inner lobes, which are provided in Table~\ref{tab:lobe-spectral-age}. From the estimates, we observe that the inner lobes have similar spectral ages, whereas the two outer lobes exhibit different spectral ages, although the spectral age differences are within the uncertainties. The northern outer lobe (NO) observed to be older than the southern outer lobe (SO).

\begin{table*}
    \centering
    \caption{Spectral age estimates of J2349-0003. The spectral indices, $\alpha$ in column 3 are estimated by fitting straight lines in the spectra given in Fig.~\ref{fig:lobe_spectrum}. $\mathrm{\tau}$ and $\mathrm{\tau^{'}}$ are the spectral ages of the radio components estimated using $\mathrm{B_{eq}}$ and $\mathrm{B_{eq}^{'}}$ respectively.}
    \label{tab:lobe-spectral-age}
    \begin{tabular}{lccccccc}
    \hline
    Region & depth & $\mathrm{\alpha}$ & $\mathrm{\nu_{b}}$ & $\mathrm{B_{eq}}$ & $\mathrm{B_{eq}^{'}}$ & $\mathrm{\tau}$ & $\mathrm{\tau^{'}}$\\
        & (kpc) & & (MHz) & $\mathrm{\mu}$G & $\mathrm{\mu}$G & Myr & Myr \\
        (1) & (2) & (3) & (4) & (5) & (6) & (7) & (8)\\
	\hline
        NO & 217.3$\pm$59.7 & -1.07$\pm$0.09 & 648 & 0.82$\pm$0.15 & 1.77$\pm$0.63 & 79.15$\pm$15.57 & 103.90$\pm$19.01 \\
        SO & 175.2$\pm$48.1 & -0.53$\pm$0.07 & 648 & 0.60$\pm$0.09 & 0.69$\pm$0.25 & 68.51$\pm$13.42 & 73.03$\pm$18.02 \\
        NI & 274.0$\pm$75.3 & -0.95$\pm$0.07 & 648 & 0.71$\pm$0.11 & 1.36$\pm$0.40 & 73.96$\pm$14.37 & 96.49$\pm$19.12 \\
        SI & 254.6$\pm$70 & -0.78$\pm$0.08 & 648 & 0.67$\pm$0.11 & 1.07$\pm$0.39 & 72.21$\pm$14.24 & 88.21$\pm$20.11 \\
    \hline
    \end{tabular}
\end{table*}

\subsection{Symmetry parameters}
We also carried out studies on various symmetry parameters of J2349-0003. Since the radio map shows asymmetries in terms of lobe length, lobe flux and lobe alignment, a detailed study in that aspect is carried out. In radio AGNs, jets are generally emitted symmetrically in opposite directions, resulting in the formation of symmetric structures \citep{2024dabhade}. Any asymmetry such as differences in lobe length, flux density, and orientation can be attributed to environmental factors. Since the radio map shows such asymmetries, a detailed study of the various symmetry parameters of J2349-000 are also carried out, as detailed below.

\subsubsection{Arm-length ratio and Lobe-flux density ratio}
\begin{figure}[ht!]
    \includegraphics[width=\columnwidth]{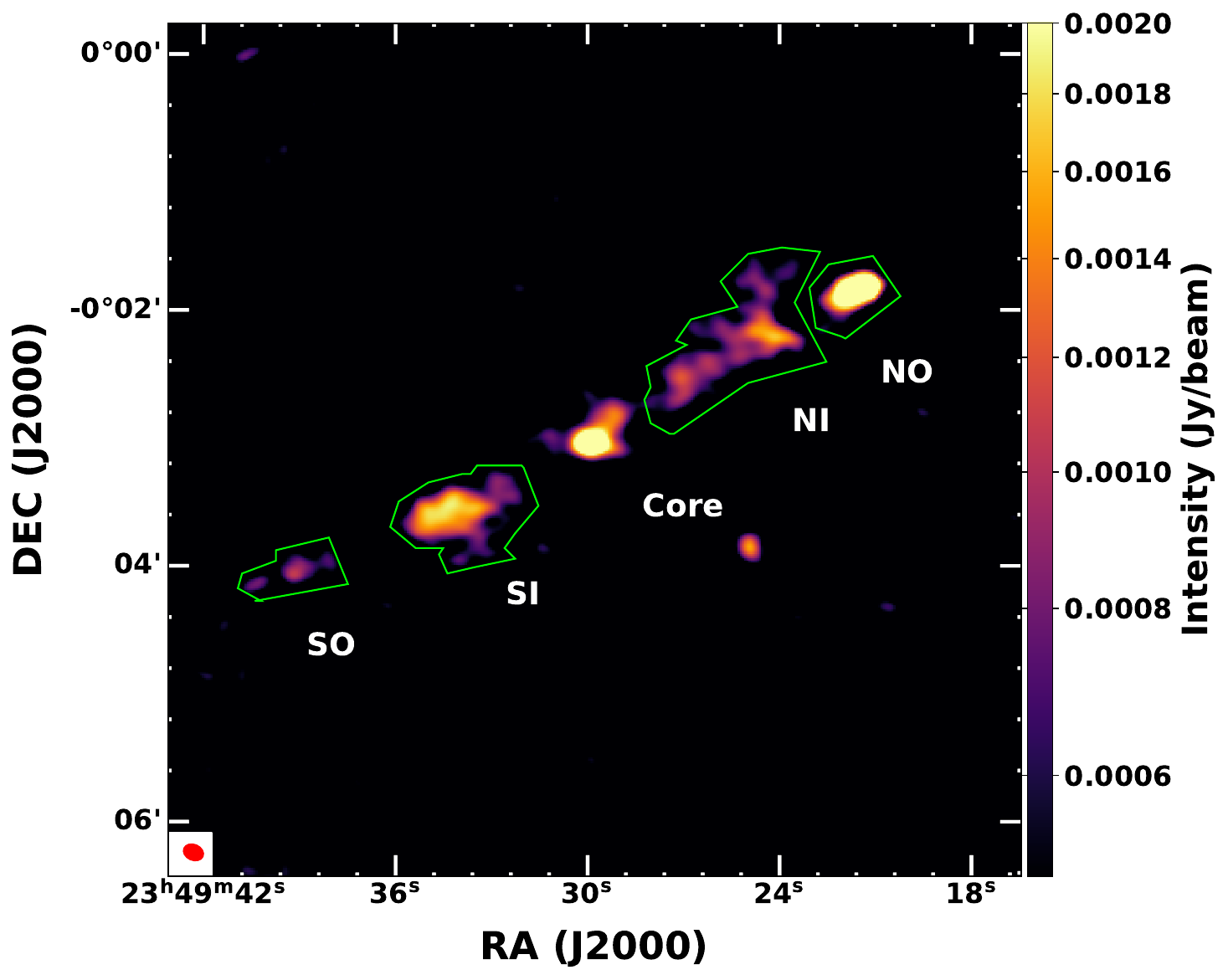}
    \caption{uMGRT Band 3 radio map. The green contours shows the regions used for extracting lobe flux densities, are overlaid over the image.}
    \label{fig:region-for-lobe-flux-ratio}
\end{figure}

\begin{figure*}
    \includegraphics[width=\textwidth]{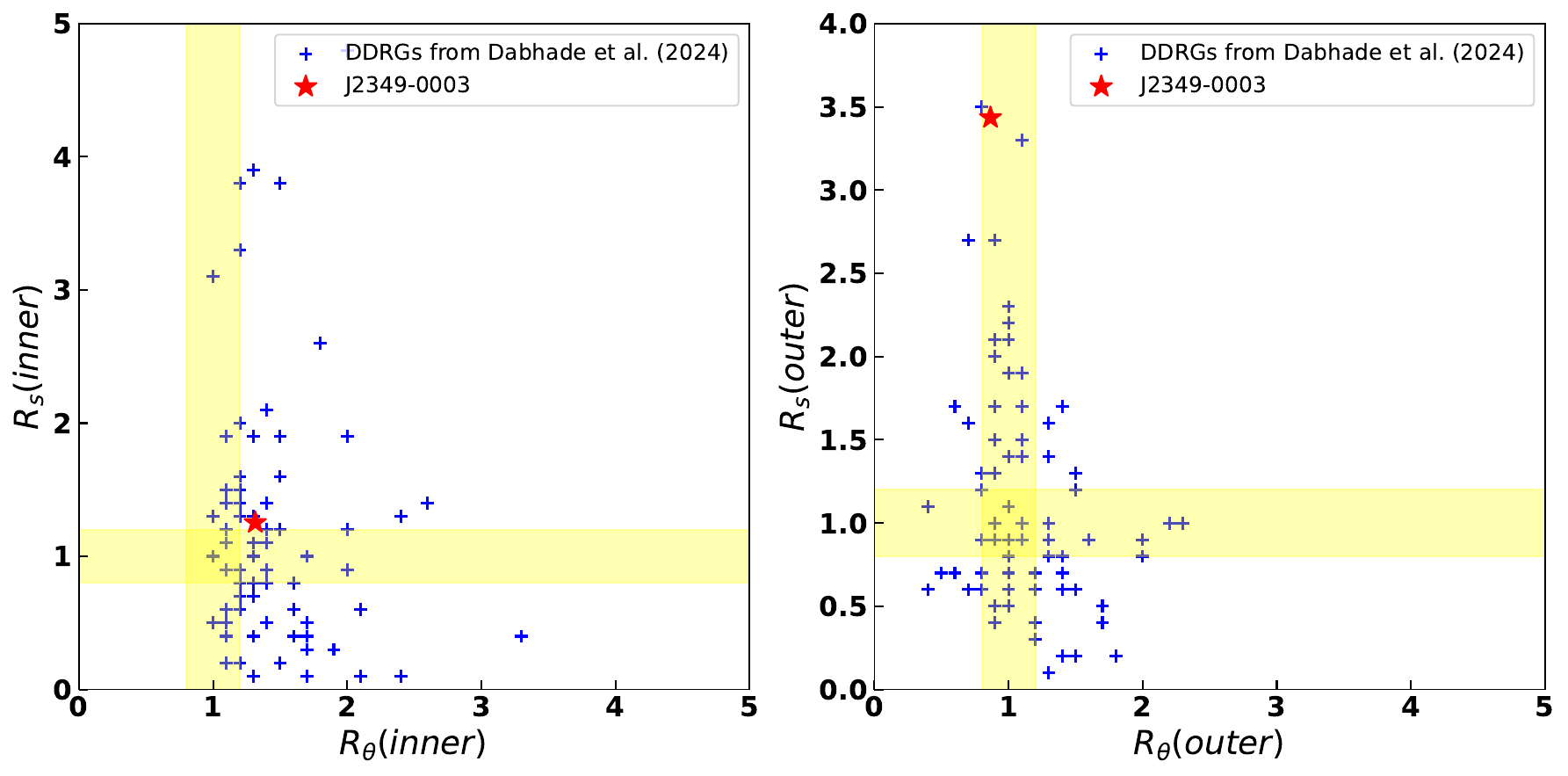}
    \caption{Left panel: Arm-length v/s Flux density ratios for the inner lobes. Right panel: Arm-length v/s Flux density ratios for the outer lobes. The blue \lq{+}\rq\ represents the data from \citet{2024dabhade}. The red star corresponds to the observations from J2349-0003. The shaded regions correspond to ratios in the range 0.8–1.2, indicating symmetry.}
    \label{fig:arm-length-and-flux-density-ratio}
\end{figure*}

Using our Band 3 radio map, we estimated the arm-length ratio (R$_{\theta}$) for both the inner and outer pairs of lobes. For the inner lobes, R$_{\theta}$ is defined as the angular size of the longer arm divided by that of the shorter arm. To estimate R$_{\theta}$ for the outer lobes, the length of the outer arm aligned with the longer inner arm is divided by the length of the opposite outer arm. This approach ensures consistency in the directional reference for the measurements. Similarly, we estimated the flux density ratios (R$_{s}$) of both pairs of lobes using our Band 3 data. The regions used for calculating the lobe flux densities are shown in Fig.~\ref{fig:region-for-lobe-flux-ratio}. The lobe length ratios and lobe flux ratios of the inner and outer lobes are provided in Table~\ref{tab:armlength-fluxdensity-ratio}. We also extracted the arm-length and flux density ratios for a sample of DDRGs from \citet{2024dabhade}, and plotted the arm-length ratio (R$_{\theta}$) versus flux density ratio (R$_{s}$), including our data points for comparison. The plot is shown in Fig.~\ref{fig:arm-length-and-flux-density-ratio}. The shaded regions in the plot correspond to ratios close to unity (0.8–1.2), indicating symmetry. A ratio of 1 for arm length suggests that the lobes are of equal size, while a flux density ratio of 1 indicates symmetric brightness between the lobes.

\begin{table}[ht]
    \centering
    \caption{Arm-length ratios and lobe-flux density ratios of the inner and outer lobes. The parameters are extracted from the Band 3 radio map of J2349-00003. NI/SI represents the ratio of parameters from northern inner lobe and southern inner lobe, and NO/SO represents the ratio of parameters from northern and southern outer lobes.}
    \label{tab:armlength-fluxdensity-ratio}
    \begin{tabular}{lcc}
    \hline
    Region & Arm-length & Lobe flux density \\
        & ratio (R$_{\theta})$ & ratio (R$_{s}$)\\
    \hline
    NI/SI & 1.31 & 1.25 \\
    NO/SO & 0.87 & 3.44 \\
    \hline
    \end{tabular}
\end{table}

From our data, we observe that the arm-length ratios for both the inner and outer lobes do not show significant deviations from symmetry. However, R{$_\theta$} for the inner lobes is slightly higher than that of the outer lobes. This discrepancy in outer R{$_\theta$} compared to the inner lobes could be influenced by environmental conditions or projection effects \citep{2024dabhade}. From the right panel of Fig.~\ref{fig:arm-length-and-flux-density-ratio}, we observe that the northern outer lobe (NO), which is closer to the core (2.45 arcmin) compared to the southern outer lobe (SO, 2.84 arcmin), is also the brighter of the two outer lobes. \citet{2024dabhade} suggest that this may indicate an interaction between the jet and the surrounding environment. All these results potentially suggest that J2349-0003 may exhibit episodic activity and has a high probability to be a DDRG. It is a giant radio source with moderately low radio power, which could indicate the presence of a sparse large-scale environment. However, the observed asymmetries imply that the source is actively interacting with its surroundings.

\subsubsection{Lobe misalignment\\}
We also checked the extent of the misalignment of the lobes of J2349-0003. Jet often restarts along the same axis as previous outbursts, with similar jet powers observed across episodes. This consistency implies that the central engine responsible for launching the jets remains relatively stable over time \citep{2023mahatma}. This implies that the radio structures produced during different episodes of jet activity are generally expected to be well-aligned. However, exceptions to this consistency have been observed in DDRGs, where significant misalignments between the various pairs of lobes can occur. Such misalignments can occur in several ways: the inner lobes may be misaligned relative to each other, the outer lobes may exhibit a similar misalignment, or the inner and outer lobes may be misaligned with respect to one another. The nature and extent of these misalignments are influenced by both the intrinsic AGN properties and the surrounding environment.

\begin{figure}[ht!]
    \includegraphics[width=\columnwidth]{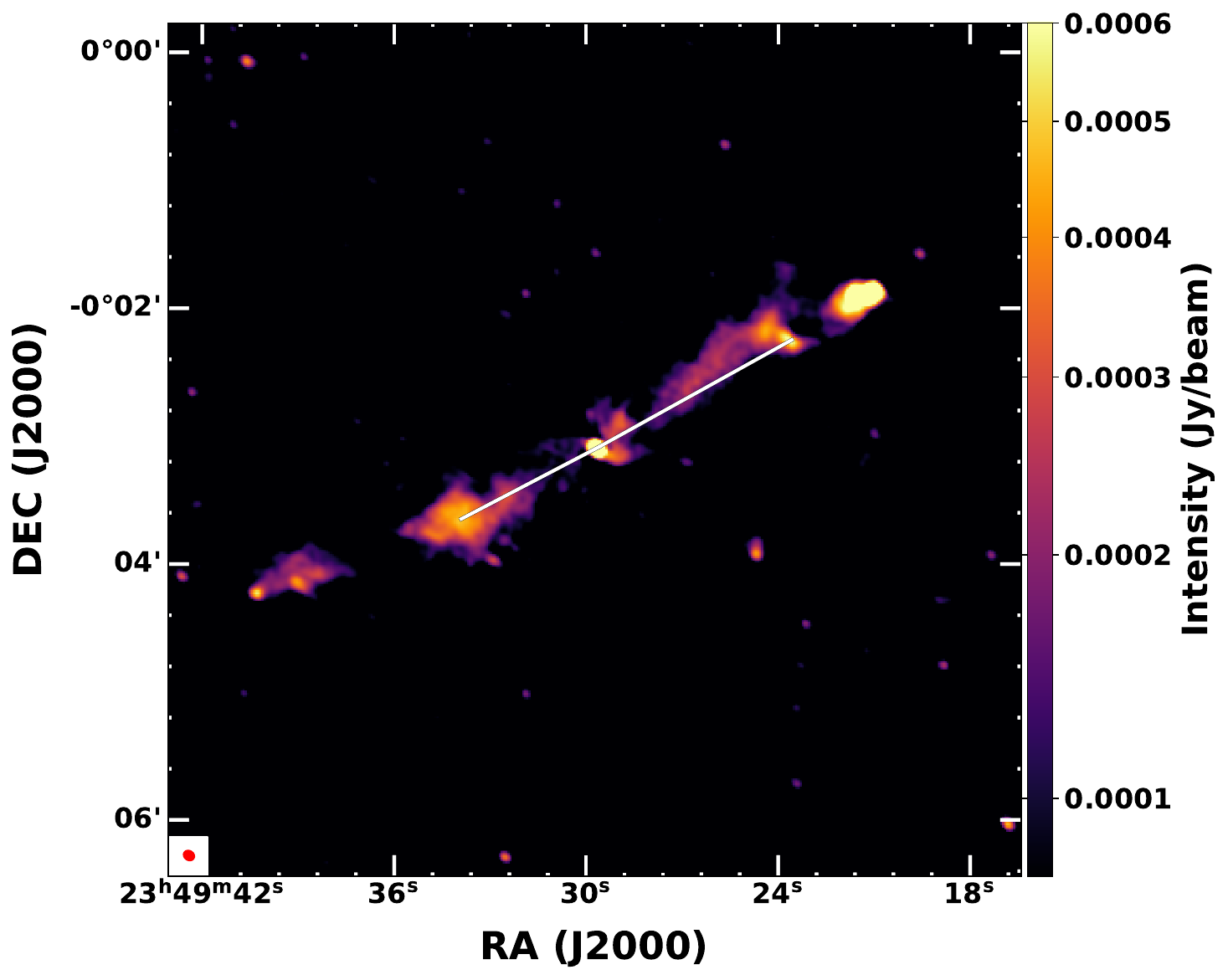}
    \includegraphics[width=\columnwidth]{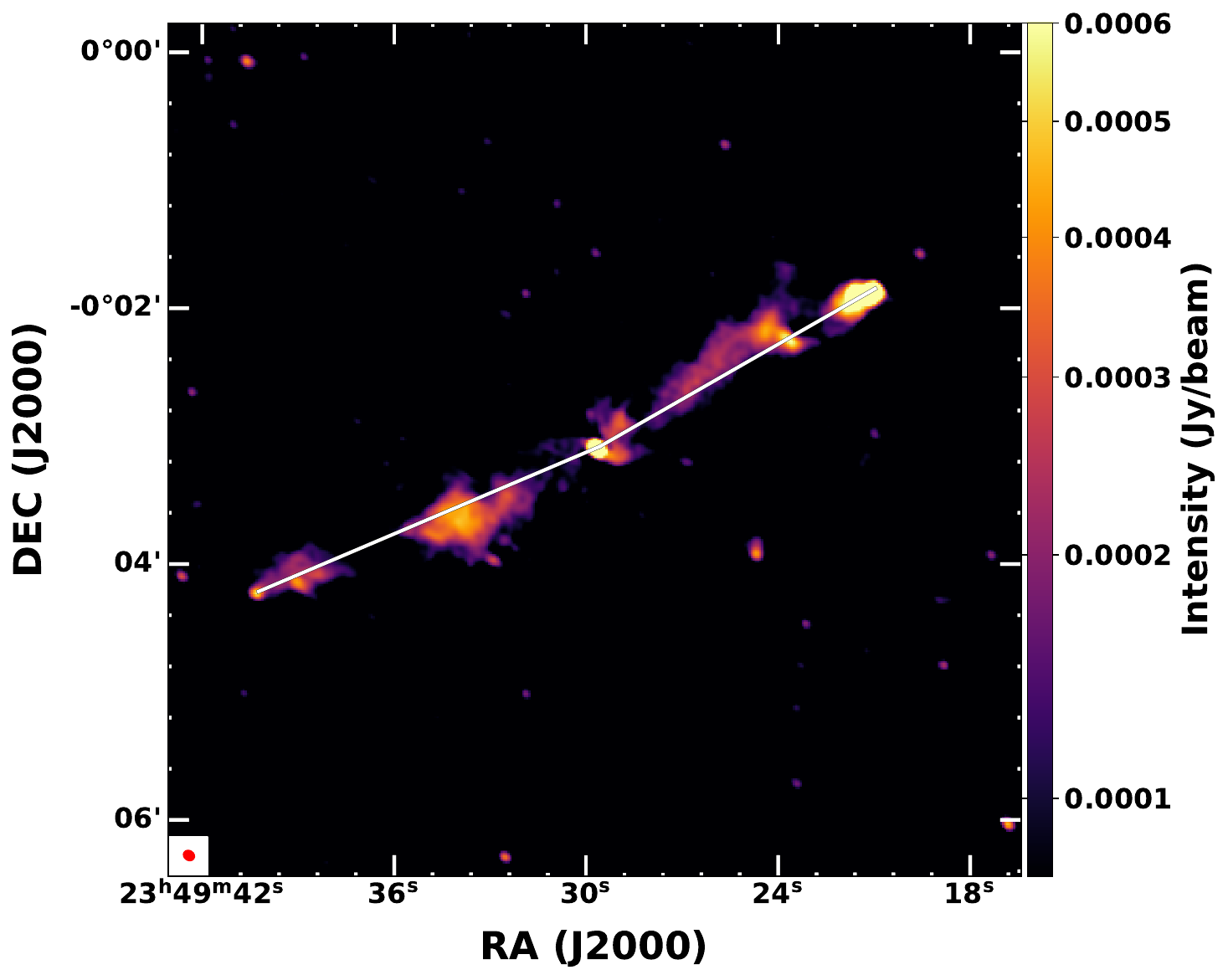}
    \caption{Top panel: The uGMRT Band 4 image with the inner lobe misalignment is shown. The white lines are drawn from the core to the peak emission region in the two inner lobes. The radio beam is shown at the bottom left corner of the figure. Bottom panel: The uGMRT Band 4 image with the outer lobe misalignment is shown. The white lines are drawn from the core to the peak emission region in the two outer lobes. The radio beam is shown at the bottom left corner of the figure.}
    \label{fig:lobe-misalignment-angle}
\end{figure}

\begin{figure}[h]
    \includegraphics[width= 3 in, height = 3 in]{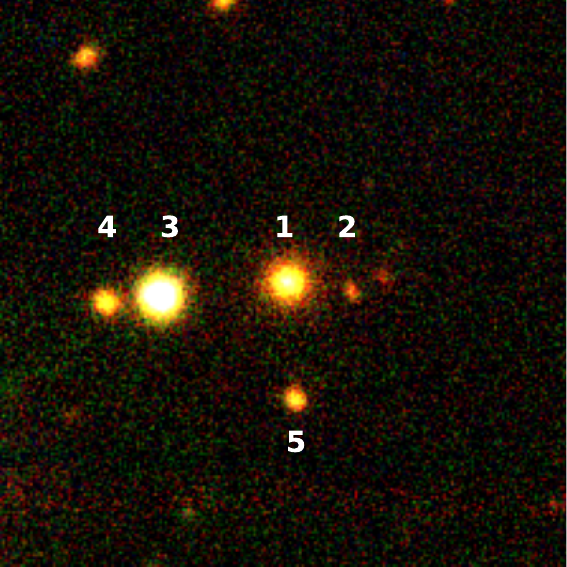}
    \caption{Pan-STARRS optical color composite image of the host galaxy of J2349-0003. The host galaxy SDSS J234929.77-000305.8 is labelled as 1. Sources 2, 4, and 5 are galaxies. Source 3 is a foreground star.}
    \label{fig:optical-image-host}
\end{figure}

In the case of J2349-0003, we estimated a misalignment angle of 1.44\textdegree\ between the two inner lobes and 6.81\textdegree\ between the two outer lobes. The plots used for estimating the misalignment angles are provided in Fig.~\ref{fig:lobe-misalignment-angle}. The observed misalignments are not extreme and could potentially result from projection effects. However, our high-resolution Band 4 images reveal that the jet emission from the core is not aligned with the axis connecting the lobes (see middle panel of Fig.~\ref{fig:DDRG-highres-images} and Fig.~\ref{fig:optical-radio-overlay}). The jets initially propagate towards the east and west, and then bend to follow the direction of the inner lobes. Therefore, the observed misalignment may be attributed to this change in jet direction near the core. Episodic AGN activity and associated misalignments are often linked to galaxy merger events. Galaxy mergers can trigger AGN activity by fueling the central SMBH, and several studies have reported evidence connecting radio AGNs with galaxy mergers \citep[e.g.,][]{2012ramos, 2022bernhard}. Recent studies also provide evidence for the role of mergers in triggering episodic activity in radio galaxies \citep[e.g.][]{2023misra}. These mergers can not only trigger AGN activity, but also cause the misalignments observed, as the merger can change the SMBH spin axis direction, eventually leading to a sudden flip in the direction of the associated jet \citep{2002merritt}. Therefore, identifying signatures of mergers is crucial in the study of misaligned DDRGs. These signatures can often be obtained from optical images. To explore this possibility for J2349-0003, we examined Pan-STARRS optical images of the host galaxy and its surroundings. The image is provided in Fig.~\ref{fig:optical-image-host}. The source labeled \lq{1}\rq\ is identified as the host galaxy at a redshift of 0.187$\pm$0.0491. A nearby galaxy, labeled \lq{2}\rq, at an angular separation of 6.5$^{\prime\prime}$ from the host, is located at a redshift of 0.399$\pm$0.1015. Source \lq{3}\rq\ is a star, while sources \lq{4}\rq\ ({\tt z}=0.345$\pm$0.0553), and \lq{5}\rq\ ({\tt z}=0.330$\pm$0.0756) are galaxies. The photometric redshift estimates of sources 1, 2, 4, and 5 indicate that they do not share similar redshifts. However, sources 4 and 5, and potentially 2, show similar redshifts within uncertainties. The redshift difference between the host galaxy and the nearby source \lq{2}\rq\ is found to be 0.212$\pm$0.1128, which rules out the possibility of their proximities to each other and any interactions between them. However, as only photometric redshift data are available, precise conclusions about possible interactions or associations cannot be drawn and it requires accurate spectroscopic redshift measurements. However we suggest the possibility of an interaction as we could see a signature in the jet direction as seen from the Band 4 radio map.

\section{Discussions}
\label{sec:discussions}
The deep, high-resolution radio observations of J2349-0003 suggest that the source may be a DDRG with a possible misalignment. Two distinct pairs of radio lobes are clearly visible in the Band 3 and Band 4 radio maps. However, the inner lobes do not exhibit any compact radio features typically associated with active jet activity. Additionally, the outer lobes are not diffuse, which is unusual for remnants from a previous epoch of AGN activity.

The radio power of J2349-0003 is estimated to be of the order of $10^{24}~\mathrm{W\,Hz}^{-1}$, indicating that it is not a highly powerful radio source capable of easily growing to giant scales. However, \citet{2022oei} suggested that high radio power is not a strict requirement for the formation of giant radio sources. Under favorable environmental conditions, or through mechanisms such as episodic jet activity, even low-power radio sources may potentially evolve into giant structures.

\begin{figure}
    \includegraphics[width=\columnwidth]{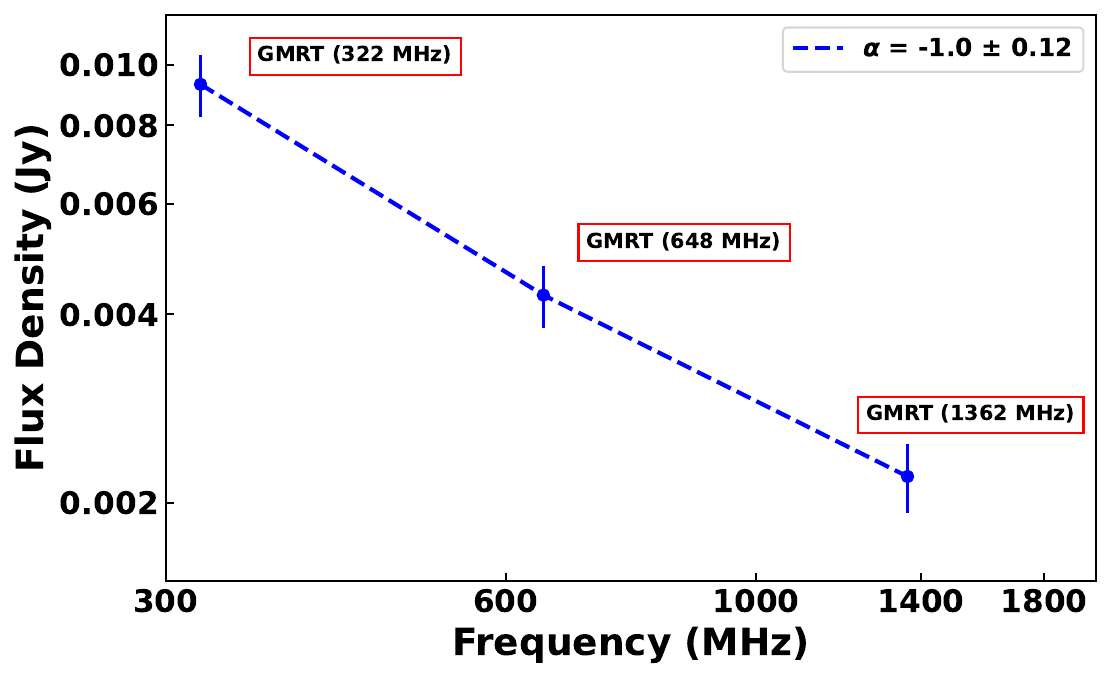}
    \caption{Core spectrum of J2349-0003. The flux densities are estimated from uGMRT Band 3, 4, and 5.}
    \label{fig:core_spectrum}
\end{figure}

The integrated spectral index of the source is estimated to be $-0.83\pm0.11$. This relatively steep spectral index suggests the presence of an aging electron population. DDRGs often retain remnant emission from previous episodes of AGN activity, which can contribute to the steepening of the integrated spectrum. 
We also estimated the spectral index of the core using the Band 3, 4, and 5 radio data, as shown in Fig.~\ref{fig:core_spectrum}. The core spectral index is found to be $-1\pm0.12$, which is relatively steep for a compact core. Generally, the compact core of an AGN is expected to exhibit a flatter radio spectrum. The steep spectral nature of the core may suggest that the core could be a compact steep spectrum source \citep[CSS;][]{1982peacock, 1998odea}, potentially containing newly formed radio doubles within the core region. Our high-resolution Band 5 radio map reveals an angular size of $\sim$7" for the core, corresponding to a projected linear size of $\sim$22 kpc, which is the typical size range for a CSS source \citep{2021odea}. However, any newly formed radio doubles within the core may be too compact to be resolved with our current resolution \citep{2015orru}.

Additionally, the core spectrum exhibits a slight concave curvature. The spectral index between Bands 3 and 4 is estimated to be –1.1$\pm$0.23, while between Bands 4 and 5, it is estimated to be –0.89$\pm$0.23. This slight flattening of the core spectrum at higher frequencies may suggest a concave spectrum, although the difference is within the uncertainties. Recent studies by \citet{2025raj} propose that such curvature could be indicative of episodic AGN activity. Therefore, the CSS characteristics of the core, combined with the observed spectral curvature, may imply that this source is a candidate triple-double radio galaxy \citep{2007brocksopp}, with the core region possibly harboring a very recent episode of jet activity \citep[e.g.,][]{2018sebastian}.

To investigate the episodic nature of J2349-0003, we also estimated the spectral ages of its radio lobes. The spectral ages of the inner pair of lobes were found to be comparable, suggesting that they may have originated from the same episode of AGN activity. In contrast, the outer lobes exhibit significantly larger spectral ages, indicative of an older population of radio plasma. However, the spectral ages of the two outer lobes are not consistent. The southern outer lobe (SO) appears to be relatively younger than the northern outer lobe (NO), with hotspot-like features visible in the Band 4 radio map (Fig.~\ref{fig:DDRG-highres-images}). This may be attributed to the injection of freshly accelerated particles or to a recently ceased jet activity \citep{2015orru}. This may indicate that the quiescent phase of the AGN between the two episodes of activity was relatively short. 

We also undertook a study on various symmetry parameters of the source. The observed morphological asymmetries in J2349-0003, including asymmetries in lobe length and flux density, suggest the possibility of environmental influence on the radio source. Importantly, the northern outer lobe (NO) is both closer to the core and brighter than the southern outer lobe (SO), indicating a possible interaction with a denser or more turbulent ambient medium. Despite these asymmetries, the inner and outer pairs of lobes exhibit relatively symmetric arm-length ratios, which potentially suggest that intrinsic jet properties remain stable across episodes. Additionally, the presence of modest misalignments in the lobes, 1.44\textdegree\ for the inner lobes and 6.81\textdegree\ for the outer lobes, suggests a change in the jet orientation over time, probably driven by dynamical processes near the core. Our high-resolution images reveal a marked deviation in the initial jet direction from the core, followed by a bending toward the lobe axis, supporting this scenario. Such misalignments are often associated with episodic AGN activity triggered by galaxy mergers, which can reorient the SMBH spin axis and lead to jet realignment. The optical image of the host galaxy environment points towards the possibility of galaxy mergers with the proximity of galaxies at short angular distances with the host. However, to confirm this, precise spectroscopic data are required.

\section{Conclusions}
\label{sec:conclusion}
We present a multi-frequency study of the candidate double-double radio galaxy (DDRG) J2349-0003, which exhibits possible lobe misalignment. Deep, high-resolution uGMRT radio maps at Bands 3 and 4 reveal the intricate radio morphology of the source, showing a pair of inner lobes and a pair of outer lobes. The core and compact components are clearly visible in the Band 5 image. The alignment of both lobe pairs with the central core strongly supports the classification of J2349-0003 as a DDRG. The projected linear size of 1.08 Mpc indicates that it qualifies as a giant radio galaxy (GRG). However, the estimated radio power of the order of $10^{24}~\mathrm{W\,Hz}^{-1}$, suggests a relatively sparse ambient environment.

The spectral age estimates of the inner and outer lobes support the occurrence of two distinct episodes of AGN activity separated by a short quiescent phase. The possible CSS nature of the core, together with the observed concave spectral curvature, suggests an ongoing or recent jet activity. These properties, when taken together, raise the possibility that J2349-0003 could even be a candidate triple-double radio galaxy. Measurements of the arm-length ratio (R$_{\theta}$) and flux density ratio (R$_{S}$) for both inner and outer lobes point towards environmental effects influencing the symmetry and morphology of the source. In addition, the presence of lobe misalignments is suggestive of dynamic phenomena, such as galaxy mergers. The orientation of the jet axis and the possible proximity of nearby galaxies to the optical host imply that merger-driven processes could be responsible for both the observed misalignment and the episodic nature of AGN activity in J2349-0003.

\section*{Acknowledgments}
We thank the anonymous reviewer for their suggestions and comments. Ruta Kale and C. H. Ishwara-Chandra acknowledges the support of the Department of Atomic Energy, Government of India, under project no. 12-R\&D-TFR-5.02-0700. VJ acknowledges the support provided by the Department of Science and Technology (DST) under the ‘Fund for Improvement of S \& T Infrastructure (FIST)’ program (SR/FST/PS-I/2022/208). VJ, and JJ also thanks the Inter-University Centre for Astronomy and Astrophysics (IUCAA), Pune, India, for the Visiting Associateship. We thank the staff of the GMRT that made these observations possible. The GMRT is run by the National Centre for Radio Astrophysics (NCRA) of the Tata Institute of Fundamental Research (TIFR). This research made use of Astropy \footnote{\url{https://www.astropy.org/}} a community-developed core Python package for Astronomy \citep{2013astropy}, and APLpy, an open-source plotting package for Python \citep{2012robitaille}.
\vspace{-1em}


\bibliography{ddrg}

\end{document}